# Microfluidics for single-cell study of antibiotic tolerance and persistence induced by nutrient limitation

Stefany Moreno-Gámez[1,2,3], Alma Dal Co[4], Simon van Vliet[5], Martin Ackermann[1,2]

[1] Institute of Biogeochemistry and Pollutant Dynamics, Department of Environmental Systems Science, ETH Zurich 8092 Zurich, Switzerland

[2] Department of Environmental Microbiology, Eawag, 8600 Dübendorf, Switzerland

[3] Groningen Institute for Evolutionary Life Sciences, University of Groningen, 9700 CC, Groningen, The Netherlands

[4] School of Engineering and Applied Sciences, Harvard University, Cambridge, MA 02139, USA

[5] Biozentrum, University Basel, Klingelbergstrasse 70, 4056 Basel, Switzerland

Correspondence: stefany.moreno@evobio.eu

**Summary**

Nutrient limitation is one of the most common triggers of antibiotic tolerance and persistence. Here, we present two microfluidic setups to study how spatial and temporal variation in nutrient availability lead to increased survival of bacteria to antibiotics. The first setup is designed to mimic the growth dynamics of bacteria in spatially structured populations (e.g. biofilms) and can be used to study how spatial gradients in nutrient availability, created by the collective metabolic activity of a population, increase antibiotic tolerance. The second setup captures the dynamics of feast-and-famine cycles that bacteria recurrently encounter in nature, and can be used to study how phenotypic heterogeneity in growth resumption after starvation increases survival of clonal bacterial populations. In both setups, the growth rates and metabolic activity of bacteria can be measured at the single-cell level. This is useful to build a mechanistic understanding of how spatiotemporal variation in nutrient availability triggers bacteria to enter phenotypic states that increase their tolerance to antibiotics.

1. **Introduction**

Bacteria that are not genetically resistant to antibiotics can enter phenotypic states that allow them to survive antibiotic concentrations that would otherwise be lethal. Increasing evidence indicates that this phenomenon can be a major cause of treatment failure in the clinics *(1-3)* and can facilitate the evolution of antibiotic resistance in the long term *(4-6)*.

Antibiotics often target cellular processes that operate when bacteria are metabolically active and dividing (e.g. the synthesis of the cell wall) *(7, 8)*. As a consequence, phenotypic states that allow bacteria to survive antibiotics are generally characterized by slower growth or complete growth arrest. Bacteria enter these states usually after being triggered by external abiotic or biotic factors. Most of these factors induce cellular stress and include starvation, acidity, and the host immune response *(9-11)*. These factors can cause a transient reduction in the growth rate in (a part of) the population, resulting in increased survival to antibiotic stress. While the term antibiotic tolerance refers to a population-wide increase in survival in the presence of antibiotics, persistence refers to the stochastic switch of only a subset of the population into protected phenotypic states *(12, 13)*. In practice it can be hard to determine which scenario is at work especially when environments vary spatially. For simplicity, we will use the term tolerance in the remainder of this chapter, since bacterial cells that can survive antibiotics without being genetically resistant are defined as tolerant regardless of whether the underlying phenomenon is tolerance or persistence *(12)*.

In nature, nutrient starvation is one of the most important factors leading to slow growing and non-growing states in bacteria. In this chapter we present two microfluidics setups to study how variation in nutrient availability in space or time can lead to increased tolerance to antibiotics (Fig. 1).

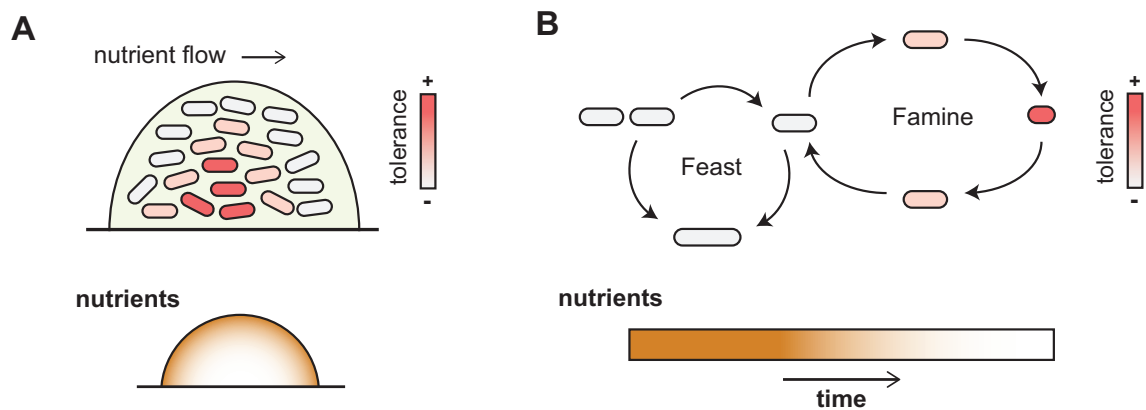

**Figure 1. Spatiotemporal regimes of variation in nutrient availability and antibiotic tolerance.** In this chapter we present microfluidic setups to study antibiotic tolerance in two scenarios that often lead to nutrient limitation in nature. **(A)** In biofilms, spatial gradients in nutrient availability are common and limit the growth and metabolic activity of cells deep inside the biofilm. As a result, these cells grow more slowly, which increases their tolerance to antibiotics. **(B)** In nature, bacteria repeatedly switch between periods of feast and famine, spending a considerable amount of time in starved, non-growing, states. These states offer protection to antibiotics because metabolic activities are minimal and cell growth is fully arrested. Some cells can remain in these non-growing states for prolonged durations even after the bulk of the population has resumed growth, which increases antibiotic survival of clonal bacterial populations.

The first microfluidic setup allows us to characterize how spatial variation in nutrient availability leads to increased antibiotic tolerance in clonal populations of bacteria *(14, 15)*. In this setup bacteria grow as two-dimensional populations inside microfluidic chambers where nutrients diffuse into the population from a single side of the chamber (Fig. 2A,B). This setup mimics biofilm growth conditions (Fig. 1A). In fact, our chambers can be seen as a two-dimensional model (i.e. a slice) of a biofilm: Although biofilms have three dimensional structures, nutrient availability in a biofilm primarily varies along a single direction away from the surface where nutrients are present. In both, natural biofilms and our growth chambers, cells take up and release nutrients and metabolites and thereby create spatial gradients in nutrient availability *(15-17)*. As a result, cells vary in their growth rate and metabolic activity depending on how far they are from the nutrient source. Under a constant flow of nutrients in the main channel of our microfluidic device, the cells in the growth chambers establish a stable gradient in nutrient availability and their growth rate decreases with the distance from the chamber opening. At this point, an antibiotic pulse can be applied to the population to study how variation in growth rate and metabolic activity increases tolerance to antibiotics in the population.

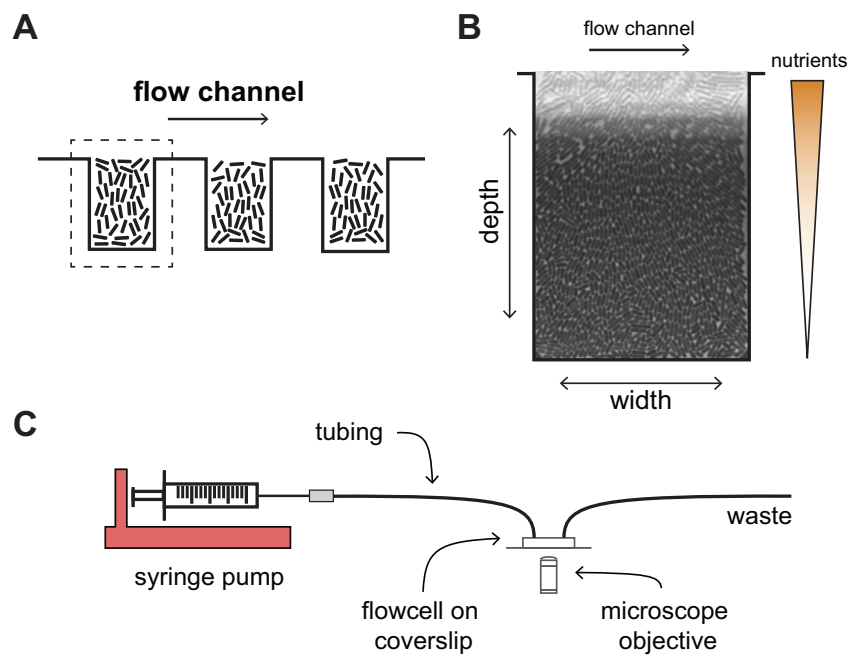

**Figure 2. Microfluidic setup to study the role of spatial heterogeneity in nutrient availability on antibiotic tolerance. (A)** Scheme illustrating the 'family' machine design of the microfluidic device. Each flow channel is ~2 cm long and there are several equally-

spaced growth chambers along the bottom and the top of the channel (only bottom side is shown). Inside a growth chamber (dotted line) bacteria grow in a densely packed two-dimensional layer. **(B)** Phase contrast image of a growth chamber from our experiments. Since nutrients diffuse into the chamber from only one side (at the top of figure), there is a spatial gradient in nutrient availability similar to those observed in biofilms. Thus, we can study the effect of phenotypic variation on antibiotic tolerance resulting from adaptation to different microenvironments (variation between cells at different depths) and phenotypic heterogeneity resulting from stochastic processes (variation between cells at same depth). The strength of the gradient can be manipulated by tuning the depth of the growth chambers or the concentration of the supplied nutrients. We used chambers that are 60 µm deep and 40 µm wide, since for these dimensions there is a strong gradient in the concentration of glucose for our experimental conditions *(14)*. The top of the image appears brighter because of imaging artifacts occurring at the chamber opening. These artifacts can be corrected when analyzing the images. **(C)** Overview of the entire setup. The growth medium is pumped into the flow cell using a syringe pump. The syringe is connected to the inlet port of the channel in the flow cell using the tubing. A short piece of thicker tubing (in grey) is used to connect the syringe needle to the tubing. Tubing also connects the outlet port of the channel to a waste container. The flow cell is placed under the microscope for continuous observation. The mould used in our experiments has eight parallel flow channels such that multiple conditions can be studied simultaneously.

The second microfluidic setup allows us to characterize how variation in nutrient availability over time leads to increased tolerance to antibiotics. In particular, our setup captures the temporal dynamics of nutrient availability characteristic of feast-and-famine regimes (Fig. 1B). These regimes are recurrently encountered by bacteria in nature and are likely to be relevant in environments like the human gut. In our setup, cells grow in a microfluidic device that is connected to a batch culture (Fig. 3). Cells in the batch culture are inoculated at a low density and gradually deplete the available nutrients until they become starved. Since bacteria in the microfluidic device are coupled to the batch culture they will follow the same transition: They first grow exponentially but then, as nutrients are depleted, their growth rate decreases until they reach full growth arrest. Bacteria in the microfluidic device remain starved for as long as they are connected to the batch culture with depleted nutrients. Then, they can be switched to fresh media and exposed to an antibiotic pulse either at the same time that nutrients are provided *(9)* or some time after this switch *(18)*. In this way, one can study how the timing of growth resumption of a cell affects the probability that the cell survives antibiotic exposure after starvation.

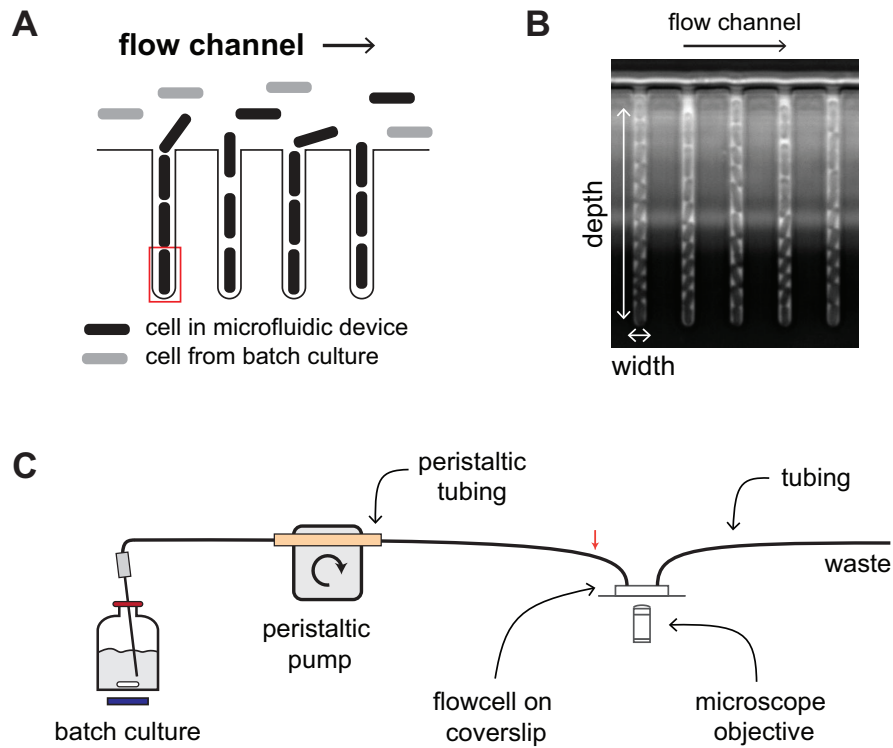

**Figure 3. Microfluidic setup to study the role of temporal heterogeneity in nutrient availability on antibiotic tolerance.** **(A)** Scheme illustrating the 'mother' machine design of the microfluidic device. Each flow channel is ~2 cm long and there are several equally-spaced growth chambers along the bottom and the top of the channel (only bottom side is shown). In order to starve bacteria in the microfluidic device, the device is connected to a batch culture. As a result, there is a continuous flow of cells from the culture passing by the main flow channel of the device. We focus our analysis only on the mother cells (red box) because these cells will always stay in the growth chambers. Also, since this setup is used to quantify phenotypic heterogeneity, using only the mother cells ensures that all cells that are analyzed are at the same distance from the opening of the growth chamber and thus experience practically identical microenvironments. **(B)** Phase contrast image of the growth chambers in our device. Chambers are 25 μm deep and 1.2-1.6 μm wide. Depending on the growth conditions, it might be advisable to modify the dimensions of the growth chambers (see **Note 5**). **(C)** Overview of the first phase of the setup where cells in the microfluidic

device transition from exponential to stationary phase. The batch culture used to starve cells in the microfluidic device is grown in a serum flask and continuously aerated using a magnetic stirrer (in blue) and a stir bar. The medium is drawn from the flask into the flow cell using a hubless needle that is connected to a peristaltic pump. Depending on the dimensions of the tubing and the needle, thick tubing (in grey) might be needed to make this connection. Once bacteria enter starvation they will remain in this phase for as long as the device is connected to the batch culture. In the second phase of our setup, starvation ends and bacteria are switched to fresh medium and exposed to the antibiotic pulse. This switch is done at a fixed location close to the inlet of the flow cell (orange arrow) and media is pumped into the flow cell using syringe pumps in the same manner as in the spatial setup (Fig. 2C).

These microfluidic setups have two features that are important to study the role of nutrient limitation on antibiotic tolerance. First, they allow us to closely mimic the conditions that lead to nutrient limitation in natural environments. In particular, in both setups nutrient availability changes as a result of the collective metabolic activity of the population, which captures the fundamental role that bacteria play in driving nutrient limitation in their own environment. Moreover, these setups are useful to disentangle different factors that may lead to antibiotic tolerance in nature. For instance, in the spatial setup one can study how antibiotic tolerance varies with the size of a biofilm by changing the dimensions of the microfluidic chambers. On the other hand, in the temporal setup, one can study how the timing and duration of antibiotic exposure relative to periods of feast and famine affects the fraction of tolerant bacteria. Both setups can also be combined to study natural scenarios where spatial as well as temporal variation in nutrient availability are relevant to understand why bacteria become tolerant to antibiotics. Second, these setups are designed to characterize the effect of nutrient limitation on antibiotic tolerance at the single-cell level. Investigating how the phenotype of a cell affects its chance to survive antibiotic exposure can provide a better understanding of the molecular mechanisms behind increased survival to antibiotics. For example, one can track cell lineages and correlate the cell size, growth rate, or division state with the probability that a cell survives antibiotic exposure. Then, one can use fluorescent reporters to integrate the previous data with information on the activity of particular metabolic or physiological functions that might affect antibiotic tolerance (e.g. synthesis of efflux pumps or toxin-antitoxin systems).

## 2. Materials

### 2.1. Bacterial strains and growth media

1. *Escherichia coli* MG1655 was used to develop and standardize both experimental setups. However, these setups can be adapted to study the effect of nutrient limitation on antibiotic tolerance in other strains of *E. coli* or other bacterial species (see **Notes 1-3**).
2. Minimal medium supplemented with a single carbon source was used for all experiments, in order to precisely control nutrient concentrations. A surfactant can be added to the medium to prevent biofilm formation (see **Note 2**). In our experiments we used M9 minimal medium supplemented with 0.01 % Tween 20 and glucose at concentrations ranging from 800 μM to 1.11 mM. Other growth media can be used if needed.

### 2.2. Preparation of microfluidics device

1. Mould with the appropriate design to produce microfluidic flow cells for each setup (see **Note 1**). The designs presented in this chapter are available at Metafluidics (see **Note 4**). Moulds can be fabricated in a cleanroom facility or ordered commercially. A brief description of the designs for each setup is given below.
    a. For spatial heterogeneity setup : Family machine design. In this design, there is a main flow channel that is ~2 cm long, 100 μm wide and 23 μm high with growth chambers attached on both sides (Fig. 2A). Growth chambers are 0.76 μm high, 40 μm wide, and 30 μm or 60 μm deep. The strength of the nutrient gradient can be controlled by varying the depth of the growth chamber or the concentration of the nutrient *(14)*.
    b. For temporal heterogeneity setup : Mother machine design. In this design, there is a main flow channel that is ~2 cm long, 200 μm wide and 21 μm high with growth chambers on both sides. These chambers are 0.93 μm high, 25 μm deep, and have a width ranging from 1.2 to 1.6 μm, so they are much narrower than the chambers of the family design (Fig. 3A) (see **Note 5**).
2. Curing agent and Polydimethylsiloxane (PDMS, Sylgard 184 Silicone Elastomer Kit, Dow Corning).
3. Vacuum desiccator with pump.

4. Plasma surface treatment system. For our experiments we used a Harrick Plasma PDC-32G-2 Plasma Cleaner.
5. Robins Instrument True-Cut Disposable Biopsy Punch 0.5 mm (see **Note 6**).
6. Scalpel.
7. Nr. 1.5 coverslips that are as large, or larger, than the microfluidic device. We used Menzel-Gläser 50 mm round coverslips.
8. Hot plate.
9. Ethanol and/or isopropanol (IPA).

## 2.3. Cell observation and microfluidic experiments

1. Inverted microscope capable of phase contrast and epi-fluorescence imaging, equipped with a temperature-controlled incubator. A high resolution objective (e.g. 100x oil objective) and autofocus system are highly recommended.
2. Microfluidic tubing with a diameter that provides a tight fitting to the inlet and outlet ports of the microfluidic flow cell. Since media will be stored and pumped from a syringe, a thicker tubing might also be also necessary for connecting the tubing to the tip of the syringe needle (see **Note 7**). The reference of the tubing used in our experiments is Teflon, inner diameter 0.3 mm, outer diameter 0.76 mm (Fisher Scientific). For the thicker tubing the reference is Microbore Tygon S54HL, inner diameter 0.76 mm, outer diameter 2.29 mm (Fisher Scientific). For the needles the reference is Sterican needles 20G, 0.9 mm x 70 mm (Braun).
3. Syringes and high precision programmable syringe pump. A multichannel pump is recommended so multiple channels from the flow cell can be used simultaneously. We used pumps from New Era Pump Systems but other pumps (e.g. from ElveFlow) are also compatible with our setups.

## 2.4. Additional equipment for setup with temporal heterogeneity in nutrient availability

1. Peristaltic pump (IPC-N24 from ISMATEC) and PharMed Ismaprene tubing with inner diameter 0.25 mm, outer diameter 2.07 mm and wall thickness 0.91 mm (VWR) to connect and pump media from the batch culture into the microfluidic device (see **Note 8)**.

2. Serum flasks with rubber lids. We use flasks with a volume of 250 mL. This can be adjusted depending on the size of the incubation box.
3. Micro magnetic stirrers with stir bars to grow and agitate the batch cultures.
4. Hubless needles to draw media from the batch cultures.
5. If it is not possible to place the serum flasks inside the incubation box from the microscope or if the temperature inside the box cannot be reliably maintained away from the objectives (check this using an infrared thermometer), a separate incubator with a small entry port for tubing and cables for the magnetic stirrers is also needed. This incubator would have to be placed next to the microscope.

## 3. Methods

### 3.1. Preparation of microfluidics devices

Next we describe the protocol to prepare a microfluidic device. Although the moulds differ for both setups, this protocol is the same in both cases.

1. Mix PDMS thoroughly with the curing agent in a 1:10 ratio. A ratio of 1.5:10 can be used to make the PDMS stiffer which can help prevent the collapse of chambers in the family machine design.
2. Pour the mix onto the mould and place into the desiccator until all air bubbles have been removed (approximately 30 min).
3. Place the mould in the oven and cure PDMS at 80 °C for 1 h.
4. Cut out the microfluidic flow cell with a scalpel and make ports for the inlet and outlet of each channel with a hole puncher. It is essential to be careful when cutting to prevent permanent damage to the mould. For silicon-based moulds make sure to minimize the amount of pressure applied on the mould to prevent breaking.
5. Before binding, wash the glass coverslip with ethanol or IPA. Then, clean the surface of the flow cell that has the imprinted features using scotch tape and rinse it with water and IPA if needed. Check that water flows through all the ports that were punched in the previous step. Finally, dry the coverslip and flow cell thoroughly with pressurized air.
6. Bind the flow cell to the glass coverslip using plasma treatment. Place the coverslip and the flow cell (with the side that has the features upwards) inside the plasma surface treatment system. Then, switch on the vacuum pump and treat both surfaces with oxygen plasma. We used a Harrick Plasma PDC-32G-2 Plasma Cleaner, at a processing pressure of 2 mbar, with 30-60 sec of high power. With other plasma treatment devices the treatment time and power need to be optimized. With the help of a tweezer, put the activated side of the flow cell on top of the activated side of the coverslip and visually confirm that binding takes place. If needed gently tap on the flow cell with the tweezers to facilitate binding (avoid tapping directly on top of the growth chambers). Afterwards, place the bonded device on top of a heated hot plate at 100 °C for 1 min. These steps might require some optimization to prevent unbinding of the flow cell from the coverslip (see **Note 9**).

### 3.2. Setup with spatial heterogeneity in nutrient availability

First, bacterial cells are loaded in the microfluidic device. Subsequently, they are grown till the growth chambers are completely full after which the nutrient gradients can be established. Afterwards, bacteria are exposed to antibiotics for a fixed window of time before finally allowing them to recover in a rich growth media to assess cell survival. These steps are explained in more detail below.

1. One day before starting the experiment, start an overnight culture for loading the microfluidic device. We start this culture from a single colony picked from a LB agar plate so the cells loaded in the device can be assumed to be genetically identical.

2. To start loading the day of the experiment, first prewet the microfluidic device with culture medium using a pipette. Then, make sure to remove as much liquid as possible by pushing air into the flow channel using an empty pipette.

3. Use centrifugation to resuspend 1 mL of the overnight culture in about 10 μL of the same growth medium (see **Note 10**).

4. Pipet ~1 uL of concentrated cells into the flow cell from the outlet of the channel. If possible, avoid that the liquid reaches all the way to the inlet.

5. Place the microfluidic device under the microscope. To load the cells inside the growth chambers, connect an empty syringe to the outlet port of the channel and bring the concentrated cell solution back and forth using a small air bubble. In this way the growth chambers will dry up sucking inside the cells in the concentrated solution (see **Note 11**). Note that only one cell is necessary within each growth chamber for the chamber to be loaded.

6. Fill a syringe with a medium that allows for rapid growth of all cells in the chamber (also the ones furthest away from the chamber opening) to quickly create a dense monolayer of cells in the chambers (see **Note 12**). In our experiments we used M9 media supplemented with 10 mM glucose.

7. Connect the tubing to the syringe needle using a short piece of thick tubing before connecting it to the inlet of the loaded channel (Fig. 2C, see **Note 13**). Make sure that the tubing is full of media without any air bubbles before making the latter connection.

8. Connect tubing to the outlet of the channel to collect the media flowing out of the flow cell in a waste container.

9. Switch the syringe pump on and let the cells grow until they fill up the growth chambers. For the strain and growth conditions we used this takes ~18 h. Set the flow rate to the final rate that will be used in the experiment (see **Note 14**). In our experiments we used 0.5 mL hr$^{-1}$.

10. Start time-lapse microscopy (see **Note 15**). The imaging protocol should be compatible with the planned data analysis and early testing of the data analysis pipeline is recommended, such that the imaging protocol can be adjusted if needed. For example, if cell tracking is planned, phase contrast images should be taken often, as automated tracking software typically requires small movement of cells between subsequent frames (we imaged every 1.75 min). A good reference is that one should be able to track cells by eye in a recorded movie without too much difficulty. Fluorescence images to measure the expression of reporter genes can be taken less often.

11. Once the growth chambers are full, switch to a medium with a low nutrient concentration such that a spatial gradient in nutrient availability arises within each growth chamber. In our experiments we used M9 media supplemented with 800 μM of glucose (see **Note 16**). This switch is done by disconnecting the tubing from the first syringe and sliding it on the syringe that contains the new media. Make sure that no air bubbles are introduced and that the thick tubing is not damaged during the switch (see **Note 17**).

12. Wait for a few hours for the system to reach a steady state. In our experiments this takes ~4 h.

13. Apply the antibiotic pulse by switching the cells to the same low nutrient medium supplemented with antibiotics for a fixed window of time (see **Note 18**). In our experiments we use streptomycin at a concentration of 50 μg mL$^{-1}$ for 3 h. The switch is done as described in step 11.

14. After the window of antibiotic exposure, switch bacteria to a medium with high nutrient concentration to determine which cells survived antibiotics. A high nutrient concentration is essential to make sure that all surviving cells have access to nutrients, irrespective of how far away they are from the chamber opening. Bacteria should remain in this condition for a long period of time to make sure that surviving cells have enough time to resume growth. In our experiments we used M9 media supplemented with 10 mM glucose for 35 h.
15. To further confirm which cells survived the antibiotic pulse, a fluorescent dye can be used in this step to quantify the membrane potential of the cells. In particular, this dye could allow one to gain information about cells that did not resume growth after the antibiotic pulse ended. Apply the fluorescent dye by switching the cells to a syringe with medium containing the dye. Then, use fluorescence imaging to determine the intracellular concentration of the dye in all cells in the population.

**3.3. Setup with temporal heterogeneity in nutrient availability**

First, cells are loaded in the microfluidic device which is connected to a flask with medium so they grow for a few hours and get adjusted to the device. Second, the flask that is connected to the device is inoculated to start the batch culture. Cells in the batch culture will start dividing and eventually deplete the nutrients at which point cells in the flow cell enter starvation. The duration of starvation is set by how long the microfluidic device remains connected to the culture with depleted nutrients. Third, bacteria in the microfluidic device are switched to fresh media. At this point the antibiotic pulse can be applied together with the switch to fresh media to reproduce batch setups like the one presented by Fridman *et al (9)*. Alternatively, cells can first be grown on fresh media before switching to antibiotics as done by Moreno-Gámez *et al (18)*. Finally, after the antibiotic pulse, bacteria are switched to fresh media to determine which cells managed to survive the pulse and further characterize the dynamics of growth resumption from starvation. These steps are explained in more detail below.

1. One day before starting the experiment, start an overnight culture for loading. We start this culture from a single colony picked from a LB agar plate so the cells loaded in the device can be assumed to be genetically identical.
2. Sterilize the serum flask in which the batch culture will be grown. The flask should have a stir bar inside and the lid should be punched with a hubless needle that goes all the way to the bottom of the flask. This needle will be used to draw the media into the flow cell.

3. On the day of the experiment, start by connecting both ends of the peristaltic tubing to the microfluidic tubing. Use tape to secure these connections. One end of the tubing will be connected to the flask with the batch culture and the other end will go into the flow cell (Fig. 3C).

4. Add the batch growth medium to the sterile serum flask (see **Note 19**). Then, connect the part of the hubless needle that sticks out from the serum flask to one end of the tubing prepared in the previous step. Depending on the size of the needle a short piece of thick tubing can be needed for this connection (Fig. 3C).

5. Bring the serum flask with the attached tubing to the incubator and place the peristaltic tubing in the peristaltic pump (see **Note 20**). Place the flask on top of the magnetic stirrer (see **Note 21**) and switch the pump on at a low rate to have liquid ready to come out at the end of the tubing that will go into the flow cell. This also helps prevent contamination.

6. Load the cells as described in section 3.2, steps 2-5. Due to the narrower chambers, loading the mother machine is harder than loading the family machine so it might be necessary to move the air bubble multiple times along the flow channel to improve loading.

7. Once the cells are loaded, connect the serum flask to the flow cell at the inlet (see **Note 13**). Then, connect a piece of tubing to the outlet of the channel to collect the medium flowing out in a waste bottle. Set the flow rate to the final rate that will be used for the experiment (in our experiments we used 0.5 mL hr$^{-1}$, see **Note 14**), and let the cells grow for a few h so they acclimate to the flow cell.

8. Start time-lapse microscopy taking images regularly in order to estimate growth rates at the single-cell level (see **Note 15**). The imaging protocol should be compatible with the planned data analysis and early testing of the data analysis pipeline is recommended, such that the imaging protocol can be adjusted if needed (see section 3.2, step 10). By recording how cells enter stationary phase one can study whether the survival of cells during starvation as well as their dynamics of growth resumption depend on their growth dynamics before starvation.

9. Inoculate bacteria from an overnight culture into the serum flask using a pipette and switch on the magnetic stirrer. Always use the same dilution factor to keep the length of the exponential phase constant. Since the lid of the serum flask has to be lifted to inoculate the cells, switch off the peristaltic pump temporarily to avoid drawing up air into the flow cell.

10. Once cells have entered stationary phase, leave the flow cell connected to the batch culture for the duration of starvation. How long cells spend in starvation will determine their lag time distribution once fresh resources are supplied and thus their tolerance to antibiotics *(18)*. One can considerably decrease the frequency of time-lapse imaging in this step since cell growth is fully arrested.

11. To end starvation, fill a syringe with the medium for growth resumption (see **Note 12**), place it on a syringe pump and connect the microfluidic tubing to the syringe needle using a piece of short thick tubing. One can choose to add antibiotics to the medium already at this step of the protocol, which would emulate previous batch setups to study antibiotic tolerance *(9)*. If so, proceed to step 13 after completing this step. To do the switch, turn off the peristaltic pump and cut the tubing at a fixed short distance (~20 cm) from the inlet to disconnect the flow cell from the batch culture (Fig. 3C). Then, connect the tubing coming from the syringe to the tubing going into the inlet of the flow cell using a piece of thick tubing. As always, avoid introducing air bubbles during the switching of the tubing (see **Note 17**). A pinch valve can be used in this step to simplify the switch from the batch culture to the fresh medium *(21)*.

12. If bacteria were not exposed to antibiotics in the previous step, prepare a syringe with the same medium used for growth resumption supplemented with antibiotics. Then, switch bacteria to this medium for a fixed window of time. This switch can be done at the same location up from the inlet where the connection was done in the previous step or at the level of the syringes as in section 3.2, step 11 (see **Note 18**). Depending on the experimental question one can vary both the time at which the pulse is applied as well as the duration of the pulse. In our experiments we applied the antibiotic pulse 8.6 h after the switch to fresh medium and the pulse lasted 80 min *(18)*. For the pulse, we used ampicillin at a concentration of 100 μg mL$^{-1}$.

13. After the window of antibiotic exposure, switch the bacteria back to a fresh medium to determine which cells survived the antibiotics and to further quantify the dynamics of growth resumption at the single cell level. Bacteria should remain in this condition for a long period of time to make sure that cells with prolonged lag times will have enough time to resume growth. In our experiments we kept observing cells for 40 h after the antibiotic pulse had ended.

14. To further confirm which cells survived the antibiotic pulse, apply a fluorescent dye to measure the membrane potential of the cells (see section 3.2, step 15).

4. **Notes**
   1. In order to adapt the setups to study other bacterial species, the dimensions of the moulds and in particular their height might have to be adjusted to the expected cell size under the planned growth conditions. It is critical that the height of the growth chambers matches closely with the expected cell size. Chambers that are too low lead to growth defects *(19)* and chambers that are too high can lead to loss of cells due to washout events and/or growth of cells in multiple layers.
   2. Some bacterial species are more prone to form biofilms that can clog the device or tubing. There are several solutions to reduce biofilm formation. First, surfactants (e.g. Tween) and Bovine serum albumin (BSA) can be added to the media at low concentration. The low concentration is important to avoid that bacteria utilize these compounds as a carbon source. Second, the microfluidic device can be pre-treated with compounds like BSA to passivate surfaces *(20)*. Third, the flow rate during the experiment can be increased. Note however that a very high flow rate can increase washout of the cells from the device. Fourth, as most cell attachment happens during loading, one can load with early exponential-phase cells (which are generally less sticky) and reduce the concentration of the cell suspension (e.g. 1 to 10x). Finally, one can use strains with loss of function mutations for genes involved in biofilm formation.
   3. It is advisable to use bacterial strains with low or no motility. Otherwise bacterial cells might swim out of the devices.
   4. Family machine design: https://metafluidics.org/devices/family-machine-to-study-spatial-gradients-in-2d-bacterial-populations/. Mother machine design: https://metafluidics.org/devices/mother-machine-to-study-feast-and-famine-dynamics/
   5. If using a mother machine device with the same dimensions as ours, make sure to use a sugar concentration that is high enough to avoid strong spatial gradients along the growth chambers. Alternatively, reduce the depth of the chambers to lengths of 10-15 µm. Washout of cells increases in shorter channels so it is important to do some testing before settling down on a particular depth. Further optimization of other dimensions of the device might be necessary to avoid clogging due to the continuous flow of cells from the batch culture *(21)*. This can be a problem especially if the batch culture reaches a high density.

6. In order to prevent leakage, it is essential that the size of the hole punchers matches the size of the tubing so there is tight fit at the inlet and outlet ports of the flow cell. The 0.5 mm punchers work well with tubing that has the dimensions specified in section 2.2.

7. The tubing can be connected directly to the syringe needle if tubing and needles with different dimensions are used. We did not use these for our experiments but they are often used in similar microfluidic setups *(20)*. The references are PTFE Tubing, 0.56 mm inner Diameter x 1.07 mm outer Diameter (Adtech) and Microlance 3, 0.55 x 25 mm syringe needles. These work well together with a 0.75 mm hole puncher.

8. Since bacteria from the batch culture are being flown into the microfluidic device, there might be biofilm formation in the peristaltic tubing. If none of the solutions proposed in **Note 2** solves this problem, try using a different type of peristaltic tubing than the one suggested here.

9. There are several modifications of the protocol that can be implemented to avoid unbinding of the flow cell from the coverslip. These include increasing the power or the length of plasma treatment, cleaning flow cells and coverslip more thoroughly with water and a solvent (e.g. IPA) and baking flow cells after binding. Another possibility is to reduce the flow rate during the experiment. However, note that a very low flow rate can increase biofilm formation.

10. In case of issues with loading the device, use a growth medium with a high carbon concentration to get a more concentrated cell solution for loading or with a composition that would make cells smaller during stationary phase. Alternatively, try loading cells that are in exponential phase instead of stationary phase.

11. It is essential to minimize the time in which cells are exposed to the air bubble to prevent cell damage or death.

12. Calculate the volume of medium needed to fill a syringe based on the flow rate that will be used and the total duration of the experiment. Always add some extra volume since liquid is often lost when setting up the experiment (e.g. when placing the syringes in the pump). It is very important to eliminate air bubbles from the medium when filling up the syringes.

13. To connect tubing to the inlet or the outlet of a channel always place the microfluidic device on a flat surface, or position an in-focus oil objective directly below the insertion side. Otherwise it is easy to break the glass coverslip. Also, make sure that the tubing goes deep into the ports.

14. When choosing the flow rate for an experiment keep in mind the trade-off underlying this choice. Low flow rates decrease the chance that the flow cell unbinds from the coverslip and prevent washouts. Also, when the flow rate is low less medium is spent during an experiment. However, low flow rates will increase the chance of biofilm formation in the tubing and in the main channel of the device.

15. Make sure to firmly secure all the parts of the setup with tape such that everything stays in place during time-lapse imaging. This includes the tubing and the slide holder with the microfluidic device.

16. Use pilot experiments to establish the relationship between the nutrient concentration and the gradient depth and to establish the required time for a stable gradient to form.

17. If an air bubble gets into the tubing when doing a media switch, make sure that the bubble rapidly passes through the microfluidic device by temporarily increasing the flow rate. Alternatively, if the tubing is long enough, one can cut off the part that contains the air bubble and re-attach it to the syringe.

18. There is a delay between the time at which the media is switched at the syringe and the time at which cells in the flow cell are exposed to the new media. The duration of this delay depends on the length of the tubing and the flow rate and is essential to take into account when characterizing the response of cells to an antibiotic pulse. This delay can be measured in a control experiment where a switch is done between a normal medium and a medium containing a fluorescent dye or fluorescent beads. A fluorescent dye or fluorescent beads can also be added to the medium containing the antibiotics to have an internal control, as long as they do not interfere with the fluorescent reporters that are used.

19. Calculate the volume of medium that needs to be added to the serum flask based on the flow rate and the total duration of the periods of exponential growth and starvation. To avoid drawing up air bubbles into the microfluidic device a 250 mL flask should always have a volume of ~20 mL left after starvation is over. Once the medium is added, make sure that the tip of the hubless needle reaches all the way to the bottom of the flask.

20. If possible, place the peristaltic pump inside the incubation box to keep the cells at a constant temperature while they travel to the flow cell. Otherwise, maximize the amount of tubing inside the incubation box and measure the temperature of the tubing outside to make sure deviations from the growth temperature are minor.

21. Since the culture inside the flask will be agitated, make sure that the flask will not slip from the top of the magnetic stirrer. Putting a piece of parafilm underneath the flask is useful to increase friction. Alternatively, consider building a supportive structure to keep the flask in place. Once the magnetic stirrer is switched on, make sure that the stir bar does not bump into the needle.

**Acknowledgements**

We thank Daniel J. Kiviet for developing large parts of the protocols discussed in this chapter. S.M.-G. and M.A. were supported by Eawag and ETH Zurich and by grants nr. 31003A_149267 and 31003A_169978 from the Swiss National Science Foundation. S.M.-G.

was also supported by Starting Independent Researcher Grant 309555 of the European Research Council and a Vidi fellowship (864.11.012) of the Netherlands Organization for Scientific Research. A.D.C. is supported by Harvard University and the Materials Research Science and Engineering Center (DMR-1420570).